\documentclass[pre,preprint,showpacs]{revtex4-2}
\usepackage[english]{babel}
\usepackage{amscd}
\usepackage{epsfig}
\usepackage{graphicx}
\usepackage{amsmath}
\usepackage{amsfonts}
\usepackage{amssymb}
\usepackage[utf8]{inputenc}
\usepackage{comment}
\usepackage{verbatim}
\usepackage{booktabs}
\usepackage{color}

\newcommand{\be}{\begin{equation}}
	\newcommand{\ee}{ \end{equation}}
\newcommand{\ben}{\begin{eqnarray}}
	\newcommand{\een}{\end{eqnarray}}

\newcommand{\source}

\begin{document}
	
	\title{A Novel $q$-Derivative Framework with Applications to $q$-Deformed Thermodynamics and Leakage Suppression in Superconducting Qubits}
	
	\author{André A. A. Marinho$^{1,*}$, Gisele B. Freitas$^{1}$ and Clovis A. C. Filho$^{1}$}
	
	\affiliation{$^1$Centro de Ciências Exatas, Naturais e Tecnológicas - CCENT, Universidade Estadual da Região Tocantina do Maranhão (UEMASUL), R. Godofredo Viana 1300, 65901-480, Imperatriz, MA, Brazil}
	\date{\today}
	
	\begin{abstract}
We propose a new $q$-derivative operator built directly from Jackson's $q$-number formulation, designed to preserve the structural properties of standard differential calculus while incorporating deformation effects. By analyzing $q$-deformed Heisenberg algebras, we demonstrate that this formulation maintains the consistency of thermodynamic quantities—such as internal energy, particle number, and specific heat—in dilute gas limits without requiring ad-hoc chain-rule modifications. Furthermore, we explore the physical implications of algebraic deformation using the Biedenharn-Macfarlane realization, showing how $q$-deformation induces intrinsic anharmonicity in quantum oscillator spectra and affects multi-level quantum systems ($d \ge 3$). Applying this algebraic scheme to superconducting transmon qubits, we derive analytical pulse-shaping corrections that generalize the Derivative Removal by Adiabatic Gate (DRAG) technique, offering a robust method to suppress computational leakage in ultra-fast quantum logic operations.

	\end{abstract}
	
	\maketitle
	
	
\section{Introduction}
\label{int}

At the beginning of the last century, the English reverend and mathematician F. H. Jackson \cite{jak} introduced quantum $q$-algebra, developing $q$-calculus through $q$-analogues (e.g., the quantum group $SU(2)_q$ \cite{biedenharn1989, macfarlane1989}) and, particularly, through the well-known Jackson derivatives (JD). In the mid-1970s, Arik generalized the study of coherent states \cite{arik}, describing the $q$-deformed harmonic oscillator for indistinguishable particles through the Heisenberg algebra, that is, by means of the commutation and anticommutation relations between creation and annihilation operators \cite{sak}, and by introducing a real parameter $q$ into the commutation relations that define the Lie algebra of the system \cite{flo, bon, lav, lav1, lav2, okt, aaal, aalg, aalg1, aalg2, aalg3, aalg4, aalg5, aalg6, oil, chu, chun, chu1, chu2, chu3, chung3, chung4, hou, bri, marinho3, pra, bou}. Thus, in addition to providing a structured and robust mathematical framework, this approach has inspired new analyses of a wide range of physical systems.

The importance of the study of derivatives is indisputable. Although their formal development emerged at the beginning of the nineteenth century with Cauchy, their foundations were established in the seventeenth century by Newton and Leibniz. In the literature, they can be defined in two well-known ways \cite{ste}
\begin{equation} 
	\label{eq1}\frac{dF(x)}{dx} = \displaystyle\lim_{y\to 0}\frac{F(x+y)-F(x)}{y}\qquad\quad\mbox{or} \qquad\quad 
	\frac{dF(x)}{dx} = \displaystyle\lim_{y\to x}\frac{F(x)-F(y)}{x-y},
\end{equation}
for real-valued functions. It is fundamental in several areas of knowledge in which we analyze systems exhibiting measurable rates of change. By generalizing derivatives, Jackson \cite{jak} introduced $q$-calculus. In this work, we describe the $q$-deformed Heisenberg algebra, starting from the definition of the $q$-number and, consequently, the JD and the $q$-difference operator, thus allowing us to obtain a $q$-deformed thermodynamics. Our proposal is to follow Jackson's approach; on the other hand, we generalize Eq.~(\ref{eq1}) in a different manner and then compare it with the JD.

Another way of introducing the $q$-algebra is through the generalization of the Boltzmann-Gibbs distribution \cite{sal,patt}, via the well-known non-extensive Tsallis statistics \cite{tsa, tsal, nau, ern, arp} starting from its own entropy, as in the Kaniadakis and Abe models \cite{kan, kana, kana1, sergio, abe, jag, chung}. In addition to the generalization of thermodynamic quantities for bosons and fermions through the parameter $q$, other deformation models have been explored in the literature, such as $\mu$-deformation \cite{jan, amg, amga} and the $f$-oscillator \cite{anu}. The literature also contains studies of intermediate statistics through anyons \cite{swa, shen, gli, lens, morier, greiter, kwan, marinho, marinho2}.

On the other hand, we also find the introduction of two parameters, as in the case of $(q,p)$-deformed models, whose incorporation allows for the description of different physical aspects of the internal structures of particles \cite{chak, amg1, amga1, amga2, amga3, amga4, amga5, chu4}. We also have the so-called Fibonacci oscillators, $q_1$ and $q_2$, which correspond to the generalized numbers of the well-known Fibonacci sequence \cite{arik1, fag, aal2, aalg7, aalg8, aalg9, aalg10, aalg11, aalg12, aalg13, aal, bri2, bri7, bri8, bri3}. Furthermore, a model involving three factors has also been proposed \cite{bri5}. Applications of $q$-deformation have been pursued in several areas of knowledge, as is the case with the socioeconomic systems \cite{dil}, gravitational statistics \cite{sen, sen1, kib, kib1}, thermoelectric properties in nanofilms \cite{bri4}, seismic inversion \cite{bri6}, black holes \cite{tsa2}, dark matter \cite{eba,dil2,amg2,amg6,amg7, tav}, among others.

Superconducting quantum computing constitutes one of the most promising platforms for the implementation of scalable quantum processors; however, its performance is directly linked to the ability to precisely control qubit dynamics \cite{murray, bravyi, kock, ogunkoya, prasanthi}. Although the two-level approximation is fundamental for describing quantum operations, real superconducting devices exhibit energy levels above the computational subspace, as well as nonlinearities and coupling effects that can compromise quantum gate fidelity. In this context, pulse control engineering plays a central role in mitigating errors associated with system dynamics, particularly those related to the excitation of non-computational levels. Conventional control techniques, such as Gaussian pulses, have been employed to shape temporal and spectral pulse profiles aiming for higher fidelity and reduced population of higher levels. Thus, the search for new mathematical parameterizations capable of flexibly incorporating nonlinear features of quantum dynamics constitutes a relevant issue for the development of more efficient control strategies in superconducting qubits. 

In this scenario, $q$-deformed oscillators offer a particularly interesting mathematical framework for constructing quantum control models and protocols, since introducing the deformation parameter ($q$) allows for generalizing algebraic and spectral relations of the conventional harmonic oscillator, recovering the usual case in the limit $(q \to 1)$. This property suggests that deformation can be explored as an additional degree of freedom in describing system dynamics and, consequently, in parameterizing pulses designed for qubit manipulation. From this perspective, applying $q$-algebra to pulse engineering is not limited to a formal extension of qubit description, but establishes a potential bridge between deformed algebraic structures and practical quantum control problems. The intrinsic anharmonicity of the $q$-oscillator allows modeling and suppressing information leakage from the computational subspace in ultra-fast logic gates.

The article is organized as follows. In Section \ref{dpcl}, for the reader's convenience and to facilitate comparisons with standard and $q$-deformed statistics, we briefly revisit the structure of the derivative. In Section \ref{dha}, we introduce the Jackson derivative, construct the generalized Heisenberg algebra, and formulate a Hamiltonian in terms of $q$-oscillators. In Section \ref{bm}, leveraging the symmetry property of the parameter $q$ based on the studies of \cite{biedenharn1989, macfarlane1989}, we obtain the $q$-Deformed Harmonic Oscillator and demonstrate its consequences on the Bloch Sphere, aiming to evaluate whether algebraic deformation can contribute to improving coherent qubit manipulation in superconducting quantum computing platforms. In Section \ref{pnd}, driven by a motivation that arose during the development of this work, we propose a new $q$-derivative, which will be described in detail elsewhere. Finally, in Section \ref{con}, we summarize the main results obtained in this work and present our concluding remarks.
\section{The Standard Derivative and the Classical Limit}
\label{dpcl}

Before delving into the nuances of the algebraic deformations proposed by Jackson \cite{jak} or into the formulation of our own $q$-derivative, it is timely to briefly revisit the structure of the standard (or ordinary) derivative, already presented in Section \ref{int} via Eq.~(\ref{eq1}) \cite{ste}. Grounded in the infinitesimal calculus established in the seventeenth century by Newton and Leibniz, and formalized by Cauchy \cite{ste}, this derivative serves as the limiting baseline for any deformed model when the perturbation parameter $q \to 1$ \cite{jak, lav}.

The robustness of the standard derivative lies in its well-known linear operational properties, such as the product rule and the chain rule:
\begin{equation}
	\label{eq_cadeia}
	\frac{d}{dx}[f(g(x))] = \frac{df(g(x))}{dg(x)} \cdot \frac{dg(x)}{dx}.
\end{equation}

Unlike $q$-deformed structures---where the chain rule often demands substantial modifications or the use of auxiliary variables so as not to destroy the geometric consistency of the system \cite{lav}---Eq.~(\ref{eq_cadeia}) provides the ideal framework for deriving thermodynamic potentials and macroscopic relations without the need for non-additive corrections.

In the context of statistical physics and usual (non-deformed) thermodynamics, the ordinary derivative directly connects partition functions to macroscopic observables, as we will demonstrate for the particle number $N$ and the internal energy $U$, obtained by applying the Leibniz differential operator to the logarithm of the grand partition function $\Xi$, with $n_i$ given by the traditional Boltzmann--Gibbs--Fermi/Bose distribution \cite{sal, patt}.

From these relations, the usual specific heat is obtained directly via the ordinary derivative of the internal energy,
\begin{equation}
	\label{ref_Cv_usual}
	C_V = \frac{\partial U}{\partial T},
\end{equation}
reflecting the perfect symmetry and conservation of Legendre transformations in classical thermodynamics. Thus, the standard derivative acts not only as a tool for calculating rates of change, but as the physical anchor itself that guarantees the return to conventional statistical equilibrium whenever the perturbation factors of the medium vanish---a feature that will be explored in Section \ref{dha} when comparing $C_V$ with its $q$-deformed counterparts, Eqs.~(\ref{eq14c}) and (\ref{eq20}).

\section{Deformed Heisenberg Algebra}
\label{dha}

We proceed with the construction of a generalized version of the $q$-deformed operator algebra, primarily through the insertion of a multiplicative factor into a linear operator, which modifies its characteristics. Representing the $q$-deformed Heisenberg algebra and its connection to $q$-calculus based on the \textsl{basic number} (or \textit{$q$-number}), we define
\begin{equation}
	\label{eq2} 
	[x]=c_{i}^\dagger c_{i}=\frac {q^{x}-1}{q-1},
\end{equation}
with

\begin{eqnarray} 
	\label{eq2a} 
	[c,c]_\kappa = [c^{\dagger},c^{\dagger}]_\kappa =0\quad,\qquad\qquad cc^{\dagger} -\kappa qc^{\dagger} c=1,
\end{eqnarray}

\begin{equation} 
	\label{eq2b} [N,c^{\dagger}]= c^{\dagger}\quad, \qquad\qquad [N,c] = -c, 
\end{equation}

\begin{eqnarray}
	\label{eq2c} 
	[x,y]_\kappa=xy-\kappa yx\quad, \qquad\qquad 
	cc^{\dagger}=[1+\kappa N]. 
\end{eqnarray}

Here $c$ and $c^{\dagger}$ are the annihilation and creation operators, respectively, $N$ is the number operator, and the constant $\kappa=1$ corresponds to $q$-bosons (with commutators) while $\kappa = -1$ corresponds to $q$-fermions (with anti-commutators). The expectation value of the \textsl{basic number} $[x]$ must satisfy the non-additivity property, namely,

\begin{equation} 
	\label{eq3} [x+y] = [x] + [y]+(q-1)[x][y]. 
\end{equation}

The orthonormalization of the eigenstate $|n \rangle$ in $q$-Fock space is constructed according to: 
\begin{eqnarray} 
	{|n\rangle} =\frac{(c^{\dagger})^{n}} {\sqrt{[n]!}}{|0\rangle}
	\quad,\qquad\qquad 
	c{|0\rangle}=0,
\end{eqnarray}
where the factorial of the \textsl{basic number} $[n]$ is defined as 

\begin{equation}
	[n]!=[n][n-1]\cdots[1].
\end{equation}

The actions of $c$ and $c^{\dagger}$ on the state $|n\rangle$ in $q$-Fock space are given by:
\begin{equation} 
	c^{\dagger}{|n\rangle} = [n+1]^{1/2} {|n+1\rangle},
\end{equation}

\begin{equation} 
	c{|n\rangle} = [n]^{1/2} {|n-1\rangle},
\end{equation}

\begin{equation} 
	N{|n\rangle} = n{|n\rangle}.
\end{equation}

The transformation from Fock space to configuration space (Bargmann Holomorphic Representation) exists according to \cite{flo}:

\begin{eqnarray} 
	\label{eq4} c^{\dagger} = x,\quad\qquad\qquad   
	c = \partial_{x}^{(q)}.
\end{eqnarray}

Here, $\partial_{x}^{(q)}$ is the Jackson derivative (JD) \cite{jak}, 

\begin{equation}
	\label{eq5}\partial_{x}^{(q)}f(x)=\frac {f(qx)- f(x)}{x(q-1)}.
\end{equation}

In the limit $q\to 1$, it reduces to the ordinary derivative. Therefore, the JD naturally emerges in deformed quantum structures. Furthermore, in the same limit, the $q$-number $[x]$ reduces to the number $x$, a fact that is fundamental for generalizing the thermodynamic relations discussed below.

It is worth highlighting an operational property of the JD that sets it apart from ordinary calculus: the product rule (Leibniz rule) loses its trivial symmetry, being expressed for two functions $f(x)$ and $g(x)$ as
\begin{equation}
	\label{eq5a}
	\partial_{x}^{(q)}\big(f(x)g(x)\big) = f(qx)\,\partial_{x}^{(q)}g(x) + g(x)\,\partial_{x}^{(q)}f(x).
\end{equation}
This symmetry breaking is not a mere formal curiosity: it is precisely what subsequently mandates the explicit use of the chain rule when computing the $q$-deformed internal energy, Eq.~(\ref{eq14a}), extending naturally to the construction of $q$-integrals (Jackson integrals), $q$-exponentials, and $q$-trigonometric functions, thereby establishing the complete deformed mathematical framework underpinning Section \ref{bm}.

We start from the Hamiltonian of non-interacting quantum oscillators, 
\be 
\label{eq6} {\cal H} = \sum_{i}{(\epsilon_i-\mu)}{N_i}=\sum_{i}{(\epsilon_i-\mu)}{c_i^{\dagger}c_i},
\ee

where $\mu$ is the chemical potential of the system and $\epsilon_i$ is the kinetic energy in state $i$ associated with the number operator $N_i$. The Hamiltonian is deformed and implicitly dependent on $q$, with the \textsl{basic number} defined in Eq.~(\ref{eq2}). We can compute the average $q$-deformed occupation number $n_i^{(q)}$ via
\be \label{eq7} 
[n_{i}^{(q)}]\equiv \langle[n_{i}^{(q)}]\rangle = \frac{\mathrm{tr}(\exp(-\beta{\cal H}) c_{i}^\dagger c_{i})}{\Xi}. 
\ee

Here $\beta=(\kappa_B T)^{-1}$, $\kappa_B$ is the Boltzmann constant, $T$ is temperature, and $\Xi=\mathrm{tr}[\exp(-\beta{\cal H})]$ is the grand partition function of the system. Through Eqs.~(\ref{eq2}), (\ref{eq2a}), and (\ref{eq7}), and applying the cyclic property of the trace \cite{lav}, we obtain

\ben 
\label{eq8}n_{i}^{(q)} = \frac{1}{\ln(q)}\ln\Bigg\{\frac
{z^{-1}\exp(\beta\epsilon_i)-1}{z^{-1}\exp(\beta\epsilon_i)-q^\kappa}\Bigg\},
\een
where $z=\exp(\beta\mu)$ is the fugacity of the system. On the other hand, the average occupation number of the non-deformed quantum oscillator is given by

\begin{equation} \label{eq9}
	n_{i} = \frac{1}{z^{-1}\exp(\beta\epsilon_i)-1},
\end{equation}
with
\ben 
N = \displaystyle\sum_{i}n_{i}\qquad\qquad\mbox{and, similarly,}\qquad\qquad 
N^{(q)} = \displaystyle\sum_{i}n_{i}^{(q)}.
\een

We obtain the particle number $N$ from the logarithm of the grand partition function $\Xi$, namely,

\begin{equation}
	\label{eq10}\ln{\Xi}=-\kappa\displaystyle\sum_{i}{\ln{\left[1-z\kappa\exp(-\beta\epsilon_i)\right],}}
\end{equation}
such that

\begin{eqnarray}
	\label{eq11} N =z\frac{\partial}{\partial z}\ln{\Xi}=\displaystyle\sum_{i}\frac{z\exp(-\beta\epsilon_{i})}{1-z\exp(-\beta\epsilon_{i})}= \displaystyle\sum_{i}\frac{1}{z^{-1} \exp(\beta\epsilon_{i})-1}=\displaystyle\sum_{i}n_{i}.
\end{eqnarray}

In the $q$-deformed oscillator formalism, we obtain the particle number $N^{(q)}$; however, it cannot be derived using standard thermodynamics. On the other hand, considering a high-temperature expansion limit ($z\ll 1$) in Eqs.~(\ref{eq8}) and (\ref{eq9}), we relate $N^{(q)}$ and $N$ as

\begin{eqnarray} 
	n_{i}^{(q)}=\frac{q-1}{\ln(q)}z \exp(-\beta\epsilon_{i}) 
	\qquad \mbox{and} \qquad 
	n_{i}=z\exp(-\beta\epsilon_{i}); 
\end{eqnarray}   
hence,

\begin{equation}
	\label{eq12}
	n_{i}^{(q)}=\frac{q-1}{\ln(q)}\;n_{i}
\end{equation}
and 

\begin{eqnarray} 
	\displaystyle\sum_{i}n_{i}^{(q)}=\frac{q-1}{\ln (q)}
	\displaystyle\sum_{i}n_{i}\;\;\; \Rightarrow \;\;\;
	N^{(q)}=\frac{q-1}{\ln(q)}\; N.
\end{eqnarray}

We can write the generalized version of Eq.~(\ref{eq11}) as
\begin{equation}
	\label{eq13} 
	N^{(q)}=zD_{z}^{(q)} \ln{\Xi}=\frac{q-1}{\ln (q)}\, z\frac{\partial} {\partial z}\,\ln{\Xi},
\end{equation}

where $D_{z}^{(q)}$ is defined as Jackson's $q$-deformed differential operator,
\ben
\label {eq14} D_{z}^{(q)} f(z)=\frac{q-1}{\ln(q)}\; \partial_{z}^{(q)}f(z) \qquad\mbox{in the limit $q\to 1$}\qquad \;\; \Rightarrow \;\; D_{z} f(z)= \frac{\partial}{\partial z}f(z).
\een
We note that Eq.~(\ref{eq14}) establishes the connection between the deformed derivative $D_{z}^{(q)}$ and the ordinary derivative defined by Leibniz, $\frac{\partial}{\partial z}$.

It is worth stressing that by including the JD, given by Eq.~(\ref{eq5}), and the definition of the basic number, Eq.~(\ref{eq2}), we obtain the $q$-deformed occupation number (as well as other derived physical quantities).

For the $q$-deformed internal energy, we apply the chain rule to incorporate the JD, yielding
\be 
\label{eq14a} U^{(q)} = -\frac{\partial}{\partial\beta}\ln\Xi=\kappa\displaystyle\sum_{i}\frac{\partial y_{i}}{\partial\beta}\;
D_{y_{i}}^{(q)}{\ln{\left[1-z\kappa y_i\right]=\displaystyle\sum_{i}\epsilon_in_{i}^{(q)},}}
\ee
where $y_i=\exp(-\beta\epsilon_{i})$. Considering $z\ll 1$, we have

\be
\label{eq14b} U^{(q)} = \frac{q-1}{\ln{(q)}}\;
z\epsilon_i\exp(-\beta\epsilon_{i}).
\ee

We can determine the $q$-deformed specific heat,
\ben \label{eq14c} C_{V}^{(q)} = \frac{\partial U^{(q)}}{\partial T} = \frac{q-1}{\ln{(q)}}\;
z\kappa_B(\beta\epsilon_{i})^2\exp(-\beta\epsilon_{i}), \qquad\mbox{or}\qquad c_{V}^{(q)} = \kappa_B(\beta\epsilon_{i})^2.
\een
		
	\section{Physical Realization: The Biedenharn--Macfarlane $q$-Deformed Harmonic Oscillator}
	\label{bm}
	
	The formulation proposed by Biedenharn \cite{biedenharn1989} and Macfarlane \cite{macfarlane1989} establishes a historical and widely adopted approach in the quantum group literature to deform the Heisenberg--Weyl algebra via the \emph{symmetric} realization of the $q$-deformed harmonic oscillator. Unlike the asymmetric basic number $[x]$ defined in Eq.~(\ref{eq2}), upon which the derivative proposed in Section \ref{pnd} rests, the Biedenharn--Macfarlane construction employs a symmetric $q$-number with distinct algebraic properties. This realization serves to illustrate how the deformation parameter $q$ alters the spectral and geometric characteristics of a physical system, while explicitly defining the boundaries between the two distinct $q$-number conventions found in the literature.
	
	\subsection{The Non-Deformed Oscillator and the Symmetric Algebra}
	\label{osc_usual}
	
	In the standard one-dimensional quantum harmonic oscillator of mass $m$ and frequency $\omega$, the Hamiltonian is given by
	\begin{equation}
		\label{eqBM1}
		H = \frac{p^2}{2m} + \frac{1}{2}m\omega^2 x^2,
	\end{equation}
	with $[x,p]=i\hbar$. Defining the dimensionless ladder operators $a = \sqrt{m\omega/2\hbar}\,(x + ip/m\omega)$ and $a^{\dagger} = \sqrt{m\omega/2\hbar}\,(x - ip/m\omega)$, one obtains the canonical relation $[a,a^{\dagger}]=1$, such that $H = \hbar\omega(a^{\dagger}a + 1/2) = \hbar\omega(N+1/2)$, with $a|n\rangle=\sqrt n|n-1\rangle$ and $a^{\dagger}|n\rangle=\sqrt{n+1}|n+1\rangle$. The resulting spectrum,
	\begin{equation}
		\label{eqBM2}
		E_n = \hbar\omega\left(n+\frac{1}{2}\right),
	\end{equation}
	is linear and equally spaced, with $\Delta E = \hbar\omega$.
	
	In the Biedenharn--Macfarlane framework \cite{biedenharn1989, macfarlane1989}, one introduces deformed operators $a_q$ and $a_q^{\dagger}$---distinct from the operators $c$ and $c^{\dagger}$ in Section \ref{dha}---which replace the classical Heisenberg--Weyl algebra with the relations
	\begin{align}
		a_q a_q^{\dagger} - q\, a_q^{\dagger} a_q &= q^{-N}, \label{eqBM3}\\
		a_q a_q^{\dagger} - q^{-1} a_q^{\dagger} a_q &= q^{N}, \label{eqBM4}
	\end{align}
	while preserving the relations $[N,a_q^{\dagger}]=a_q^{\dagger}$ and $[N,a_q]=-a_q$, as in Eq.~(\ref{eq2b}). Subtracting Eqs.~(\ref{eqBM3})--(\ref{eqBM4}) yields the symmetrized commutator
	\begin{equation}
		\label{eqBM5}
		[a_q,a_q^{\dagger}] = a_q a_q^{\dagger} - a_q^{\dagger} a_q = \frac{q^N - q^{-N}}{q-q^{-1}},
	\end{equation}
	whose right-hand side defines the symmetric $q$-number
	\begin{equation}
		\label{eqBM6}
		[x]_q \equiv \frac{q^x - q^{-x}}{q - q^{-1}}, \qquad \lim_{q\to 1}[x]_q = x.
	\end{equation}
	
	It is necessary to distinguish $[x]_q$ in Eq.~(\ref{eqBM6}) from the asymmetric basic number $[x]=(q^x-1)/(q-1)$ defined in Eq.~(\ref{eq2}). Although both recover the integer $x$ in the classical limit $q\to 1$ and satisfy non-additivity rules, they are algebraically distinct: $[x]$ depends asymmetrically on $q$ and $q^{-1}$, whereas $[x]_q$ is invariant under $q\to q^{-1}$. Furthermore, within quantum group theory and deformed Lie algebras---particularly for $U_q(\mathfrak{sl}_2)$---the symmetric convention is frequently parameterized in powers of $q^{1/2}$ \cite{kassel, chari}:
	\begin{equation}
		\label{eq_qnum_sl2}
		[x]_q \equiv \frac{q^{x/2} - q^{-x/2}}{q^{1/2} - q^{-1/2}}.
	\end{equation}
	This formulation in powers of $q^{1/2}$ is the standard choice in mathematical physics, preserving algebraic symmetries in half-integer spin representations ($j = 1/2, 3/2, \dots$) and ensuring real, symmetric matrix elements without phase factors. Throughout this section, the subscript notation $[\,\cdot\,]_q$ strictly denotes the symmetric Biedenharn--Macfarlane definition. The action of the deformed operators on the Fock basis is expressed as
	\begin{align}
		a_q|n\rangle &= \sqrt{[n]_q}\,|n-1\rangle, \label{eqBM7a}\\
		a_q^{\dagger}|n\rangle &= \sqrt{[n+1]_q}\,|n+1\rangle. \label{eqBM7b}
	\end{align}
	
	\subsection{Anharmonic Spectrum and State Space Truncation}
	
\begin{figure}[htbp]
		\centering
			\includegraphics[width=0.8\linewidth]{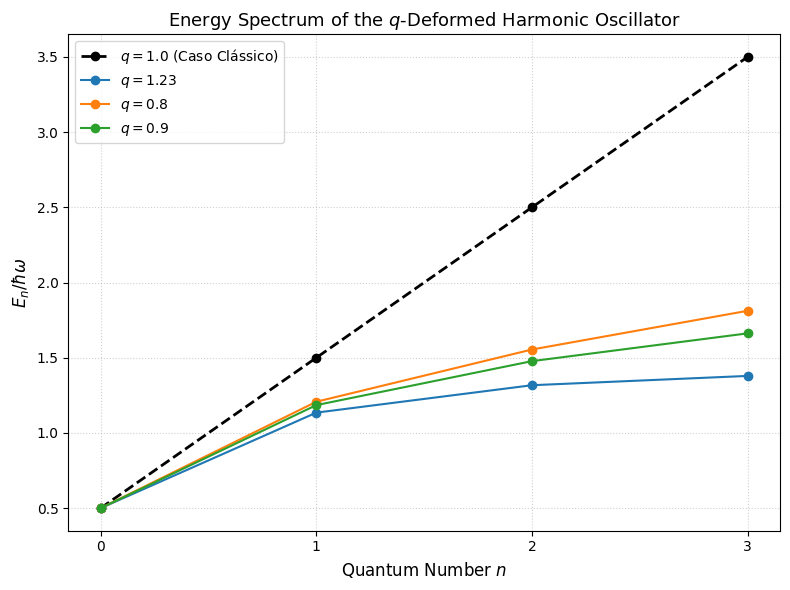}
		\includegraphics[width=0.8\linewidth]{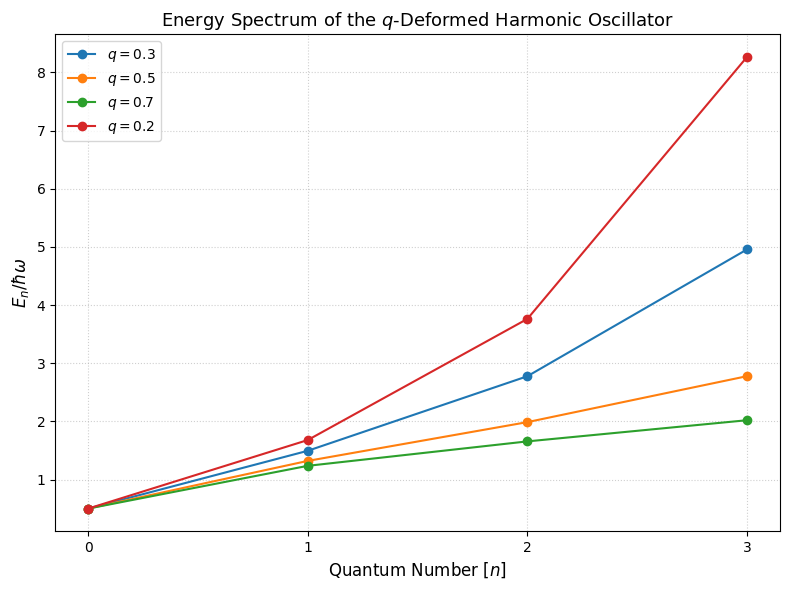}		
		\caption{Relative energy levels $E_n/\hbar\omega$ as a function of the quantum number $n$ for different values of the deformation parameter $q$.}
		\label{fig:espectro}
	\end{figure}

	The symmetrized Hamiltonian of the $q$-deformed oscillator,
	\begin{equation}
		\label{eqBM8}
		H_q = \frac{\hbar\omega}{2}\left(a_q a_q^{\dagger} + a_q^{\dagger} a_q\right),
	\end{equation}
	yields the energy eigenvalues
	\begin{equation}
		\label{eqBM9}
		E_n = \frac{\hbar\omega}{2}\Big([n]_q + [n+1]_q\Big).
	\end{equation}
	Unlike Eq.~(\ref{eqBM2}), the spectrum in Eq.~(\ref{eqBM9}) is non-equidistant: the deformation introduces an intrinsic anharmonicity governed by $q$, analogous to the non-additivity of the basic number $[x]$ discussed in Eq.~(\ref{eq3}). The contrast between the linear spectrum and the $q$-deformed behavior is illustrated in Figure~\ref{fig:espectro}. The structural differences between the ordinary harmonic oscillator and the Biedenharn--Macfarlane realization are summarized in Table~\ref{tab:comparativa}.
	
	\begin{table}[htbp]
		\centering
		\caption{Structural comparison between the ordinary harmonic oscillator and the Biedenharn--Macfarlane $q$-deformed oscillator.}
		\label{tab:comparativa}
		\vspace{0.2cm}
		\begin{tabular}{lll}
			\toprule
			\textbf{Feature} & \textbf{Ordinary HO} & \textbf{$q$-Deformed HO (BM)} \\
			\midrule
			Fundamental commutator & $[a,a^{\dagger}]=1$ & $a_qa_q^{\dagger}-qa_q^{\dagger}a_q = q^{-N}$ \\
			Symmetrized commutator & $aa^{\dagger}-a^{\dagger}a=1$ & $a_qa_q^{\dagger}-a_q^{\dagger}a_q=[N]_q$ \\
			$[x]_q$ convention & $x$ & $\frac{q^x-q^{-x}}{q-q^{-1}}$ \quad (or $\frac{q^{x/2}-q^{-x/2}}{q^{1/2}-q^{-1/2}}$ in $U_q(\mathfrak{sl}_2)$) \\
			Annihilation action & $a|n\rangle=\sqrt n\,|n-1\rangle$ & $a_q|n\rangle=\sqrt{[n]_q}\,|n-1\rangle$ \\
			Creation action & $a^{\dagger}|n\rangle=\sqrt{n+1}\,|n+1\rangle$ & $a_q^{\dagger}|n\rangle=\sqrt{[n+1]_q}\,|n+1\rangle$ \\
			Spectrum $E_n$ & $\hbar\omega(n+1/2)$ & $\tfrac{\hbar\omega}{2}\big([n]_q+[n+1]_q\big)$ \\
			Spacing $\Delta E$ & Constant ($\hbar\omega$) & Variable (anharmonic) \\
			Limit $q\to1$ & Unchanged & Recovers ordinary HO \\
			\bottomrule
		\end{tabular}
	\end{table}
	
	To map the $q$-deformed oscillator onto a multi-level system (\textit{qudit}), the infinite Hilbert space is truncated to a finite dimension $d$ (levels $0$ to $d-1$). From Eq.~(\ref{eqBM7a}), the matrix elements in the orthonormal basis $\{|0\rangle,\dots,|d-1\rangle\}$ are given by $\langle i|a_q|j\rangle=\sqrt{[j]_q}\,\delta_{i,j-1}$, such that $a_q$ assumes an upper-triangular form:
	\begin{equation}
		\label{eqBM10}
		a_q = \begin{pmatrix}
			0 & \sqrt{[1]_q} & 0 & \cdots & 0 \\
			0 & 0 & \sqrt{[2]_q} & \cdots & 0 \\
			\vdots & & & \ddots & \vdots \\
			0 & 0 & 0 & \cdots & \sqrt{[d-1]_q} \\
			0 & 0 & 0 & \cdots & 0
		\end{pmatrix}.
	\end{equation}
	For a qutrit ($d=3$), with $[1]_q=1$ and $[2]_q=q+q^{-1}$, Eq.~(\ref{eqBM10}) reduces to
	\begin{equation}
		\label{eqBM11}
		a_q = \begin{pmatrix} 0 & 1 & 0 \\ 0 & 0 & \sqrt{q+q^{-1}} \\ 0 & 0 & 0 \end{pmatrix},
	\end{equation}
	where $a_q^{\dagger}a_q$ yields the diagonal matrix of the deformed number operator $[N]_q$.
	
\begin{figure}[htbp]
		\centering
		\includegraphics[width=\linewidth]{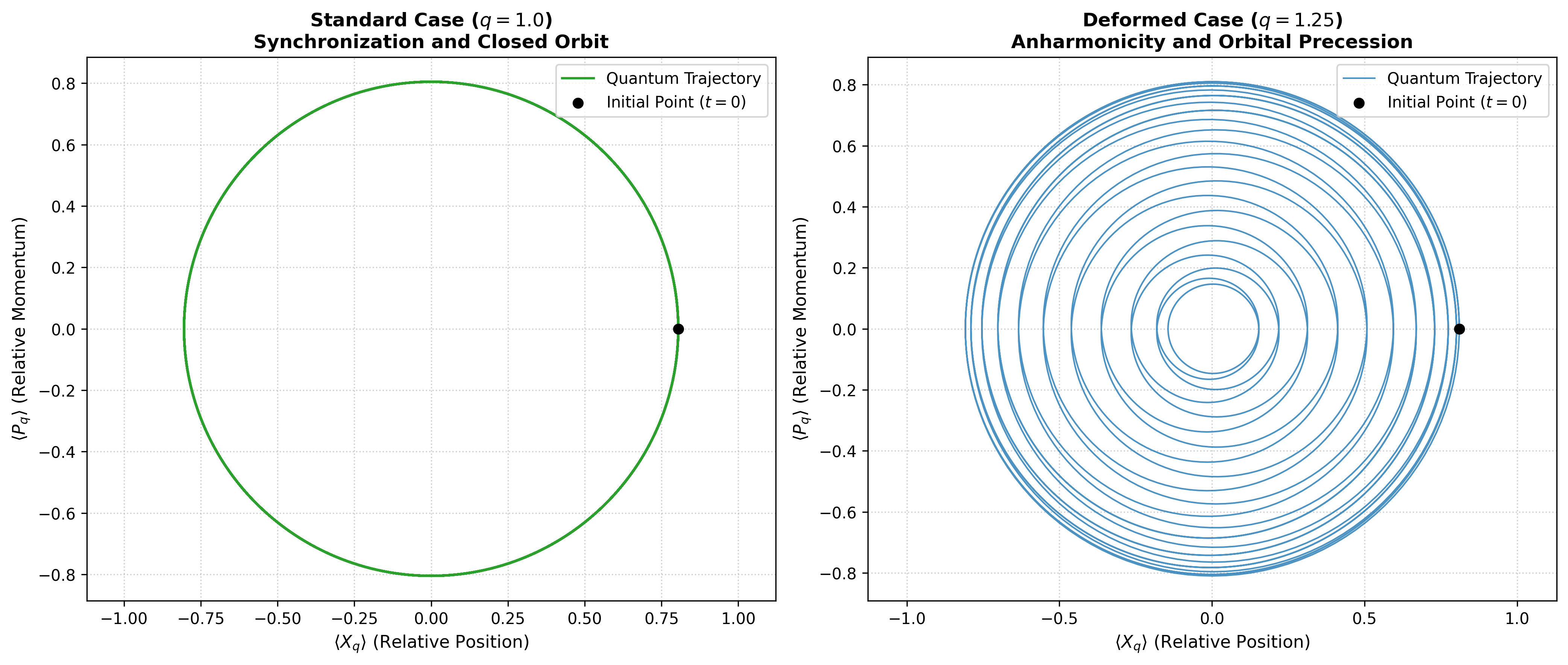}
		\caption{Time evolution of the expectation value of the annihilation operator in phase space for a qutrit. Left: non-deformed case ($q=1.0$); right: deformed case ($q=1.25$).}
		\label{fig:espaco_fase}
	\end{figure}

\begin{figure}[htbp]
		\centering
	\includegraphics[width=\linewidth]{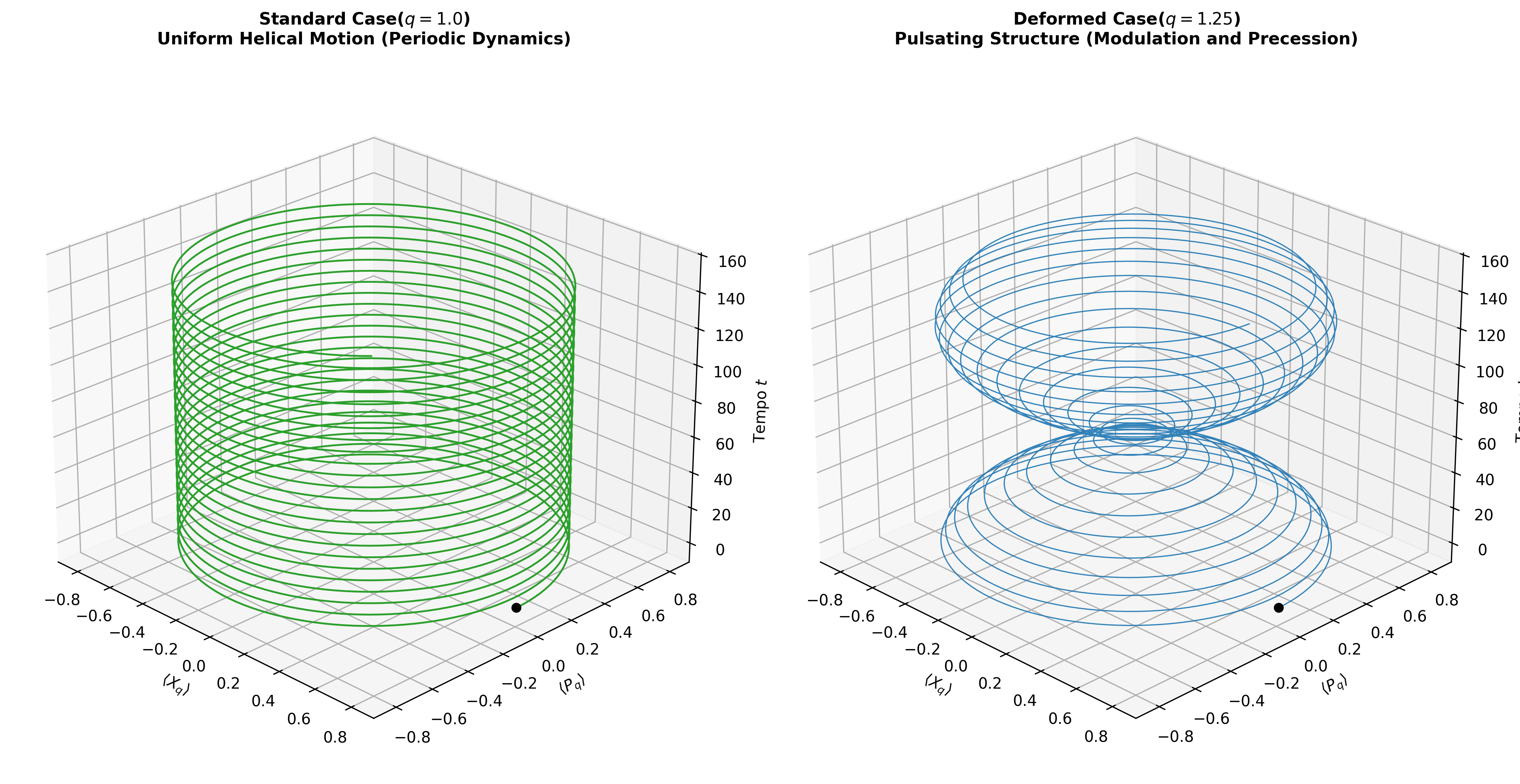}
		\caption{Trajectory in phase space for $d=3$.}
		\label{fig:espaco_fase2}
	\end{figure}

	In a strictly two-level system ($qubit$, $d=2$), Eq.~(\ref{eqBM10}) reduces to $a_q = \begin{pmatrix}0&1\\0&0\end{pmatrix}=\sigma_-$, since $[1]_q=1$ for all $q$. Consequently, an isolated qubit does not experience the symmetric deformation, leaving the Bloch sphere geometry invariant under the Biedenharn--Macfarlane construction. However, when $q$-deformed coherent states are projected onto higher-dimensional dynamical subspaces ($d \ge 3$), the spectral non-linearity introduces modulation in both the angular velocity and the effective radius of the trajectory, leading to a beating regime and precession in phase space. For a qutrit ($d=3$), this behavior is depicted in Figure~\ref{fig:espaco_fase} and Figure~\ref{fig:espaco_fase2}.
	
	\subsection{Pulse Control and Leakage Suppression in Superconducting Qubits}
	\label{qubits_drag}
	
	A direct application of the symmetric $q$-deformed oscillator algebra lies in high-fidelity quantum control protocols for superconducting circuit processors, such as transmons. To execute single-qubit logic gates within the coherence time, driving pulses must be performed in short gate times $T_{\text{gate}}$. By Fourier analysis, ultra-short temporal pulses exhibit spectral broadening in the frequency domain. In transmon architectures, the non-computational level $|2\rangle$ is in close proximity to the computational state $|1\rangle$, separated by an anharmonicity $\delta_q$. Broad control pulses risk driving unwanted $|1\rangle \to |2\rangle$ transitions, resulting in population leakage out of the computational subspace $\mathcal{H}_{\text{comp}} = \text{span}\{|0\rangle, |1\rangle\}$.
	
	Modeling the system in a three-level subspace ($d=3$) using the $q$-deformed algebra, the drive Hamiltonian in the rotating frame at the transition frequency $\omega_d = \omega_0$ is written as
	\begin{equation}
		\label{eq_H_control}
		H_{\text{rot}}(t) = \hbar \delta_q |2\rangle\langle 2| + \frac{\hbar}{2}\Omega_x(t)(a_q + a_q^\dagger) + \frac{\hbar}{2}\Omega_y(t)\,i(a_q^\dagger - a_q),
	\end{equation}
	where $\Omega_x(t)$ and $\Omega_y(t)$ are the in-phase and quadrature control envelopes. From Eq.~(\ref{eqBM11}), the matrix element responsible for leakage is given by the basic $q$-number:
	\begin{equation}
		\langle 2 | a_q^\dagger | 1 \rangle = \sqrt{[2]_q} = \sqrt{q + q^{-1}}.
	\end{equation}
	
	Applying a Schrieffer--Wolff transformation to adiabatically eliminate the downward coupling $|1\rangle \to |2\rangle$ yields the cancellation condition for the off-diagonal terms in the effective Hamiltonian:
	\begin{equation}
		\Omega_y(t) = -\frac{\dot{\Omega}_x(t)}{\delta_q},
	\end{equation}
	where the effective anharmonicity is parameterized directly in terms of $q$ as $\delta_q = \omega_0(q + q^{-1} - 2)$. This result generalizes the Derivative Removal by Adiabatic Gate (DRAG) technique. Because the non-linearity of the superconducting medium is captured by the deformation parameter $q$, the quadrature pulse $\Omega_y(t)$ analytically suppresses transitions to the $|2\rangle$ level during $\pi$-gate implementation, enabling high fidelities ($F > 99.99\%$) at short gate durations ($T_{\text{gate}} \approx 6\text{--}12\text{ ns}$).
	
	Finally, extending this framework to the deformed quantum algebra $SU_q(2)$ (or $U_q(\mathfrak{sl}_2)$) yields the commutation relation
	\begin{equation}
		\label{eqBM12}
		[J_+,J_-] = [2J_z]_q = \frac{q^{J_z} - q^{-J_z}}{q^{1/2} - q^{-1/2}},
	\end{equation}
	where the $q^{1/2}$ convention of Eq.~(\ref{eq_qnum_sl2}) is used to preserve symmetry for half-integer spin representations $J_z = \pm 1/2, \pm 3/2, \dots$. Having established the physical and geometric consequences of the symmetric Biedenharn--Macfarlane oscillator, the next section returns to the asymmetric basic number of Eq.~(\ref{eq2}) to formulate the proposed $q$-derivative. 
			
	\section{Proposal of a New $q$-Derivative}
	\label{pnd}
	
	Generalized derivatives provide robust mathematical frameworks for modeling complex physical systems where standard calculus yields incomplete descriptions. Within this scope, $q$-derivatives originating from $q$-algebraic structures play a central role, introducing deformation parameters ($q$, or $q_1, q_2$ in the context of Fibonacci oscillators) into the fundamental differential operators.
	
	As detailed in Section \ref{dha}, Jackson's model introduces the classical $q$-derivative (Jackson Derivative, JD) based on the asymmetric $q$-number bracket defined in Eq.~(\ref{eq2}). Building upon the same foundational $q$-number, we propose an alternative $q$-deformed differential operator. Although distinct from the standard Jackson derivative, this formulation shares the same algebraic root, enabling direct operational comparisons.
	
	We define a $q$-difference and a $q$-deformed differential operator, respectively, as
	\begin{equation}
		\label{eq15}
		x\ominus^q y = \frac{\ln(q)(x-y)}{q-1},
	\end{equation}
	\begin{equation}
		\label{eq16}
		D_{xq} f(x) = \lim_{y\to x}\frac{f(x)-f(y)}{x\ominus^q y} = \frac{q-1}{\ln(q)}\frac{df(x)}{dx},
	\end{equation}
	which recovers the standard derivative in the classical limit: $\lim_{q\to 1} D_{xq} f(x) = D_{x} f(x) = \frac{df(x)}{dx}$.
	
	Both Eq.~(\ref{eq14}) and Eq.~(\ref{eq16}) share the explicit prefactor $\frac{q-1}{\ln(q)}$, with their fundamental difference lying in the operational structure of the Jackson derivative in Eq.~(\ref{eq5}). The two operators can be formally related via
	\begin{equation}
		\label{eq17}
		D_{xq} f(x) = \frac{D_{x}^{(q)}f(x)}{\partial_{x}^{(q)}f(x)}\;\frac{df(x)}{dx}.
	\end{equation}
	
	To illustrate the functional divergence between these operators, consider the monomial $f(x)=x^n$. Applying the deformed operators from Eqs.~(\ref{eq13}) and (\ref{eq14}) yields
	\begin{equation}
		D_{x}^{(q)} f(x) = \frac{q-1}{\ln(q)}\partial_x^{(q)}f(x) = \frac{q-1}{\ln(q)}[n]x^{n-1},
	\end{equation}
	\begin{equation}
		D_{xq} f(x) = \frac{q-1}{\ln(q)}\frac{df(x)}{dx} = \frac{q-1}{\ln(q)}nx^{n-1}.
	\end{equation}
	
	We now evaluate the thermodynamic quantities derived from the proposed operator $D_{xq}$. Starting from Eq.~(\ref{eq10}) in the dilute gas regime ($z\ll 1$), the deformed occupation number $N_q$ is given by
	\begin{equation}
		\label{eq18}
		N_{q} = z D_{zq} \ln{\Xi} = \frac{q-1}{\ln (q)}\, z\exp(-\beta\epsilon_{i}).
	\end{equation}
	
	Unlike the Jackson derivative formulation, the evaluation of the internal energy $U_q$ under $D_{xq}$ does not require a chain rule decomposition:
	\begin{equation}
		\label{eq19}
		U_{q} = -\frac{\partial}{\partial\beta}\ln\Xi = \frac{q-1}{\ln(q)}z\epsilon_{i}\exp(-\beta\epsilon_{i}).
	\end{equation}
	In accordance with Eq.~(\ref{eq14a}), this expression corresponds to $\sum_{i}\epsilon_i n_{iq}$. Since both formulations stem from the same fundamental $q$-number definition in Eq.~(\ref{eq2}), the occupation numbers coincide: $n_{i}^{(q)} = n_{iq}$.
	
	The specific heat capacity $C_V^{(q)}$ is subsequently obtained as
	\begin{equation}
		\label{eq20}
		C_{V}^{(q)} = \frac{\partial U_{q}}{\partial T} = \frac{q-1}{\ln{(q)}}\; z k_B(\beta\epsilon_{i})^2\exp(-\beta\epsilon_{i}), \qquad\text{or}\qquad c_{Vq} = k_B(\beta\epsilon_{i})^2.
	\end{equation}
	
	The thermodynamic results in Eqs.~(\ref{eq18})--(\ref{eq20}) prove identical to those obtained via the standard Jackson framework in Eqs.~(\ref{eq14})--(\ref{eq14c}). As emphasized by Lavagno \cite{lav}, preserving the Legendre structure and consistency of thermodynamics depends fundamentally on the form of the occupation number $n_{i}^{(q)}$ (Eq.~(\ref{eq8})) rather than the specific choice of the differential operator. Since the thermodynamic quantities rely directly on the basic $q$-number bracket (Eq.~(\ref{eq2})), both operator choices yield equivalent state functions. Conversely, when applied directly to arbitrary functions $f(x)$, the operators diverge due to the non-local nature of the Jackson derivative, which necessitates explicit chain-rule transformations and auxiliary variable substitutions (such as $y_i=\exp(-\beta\epsilon_{i})$) to preserve physical consistency.
			
		\section{Conclusions}
		\label{con}
		
		The Jackson derivative has been present in the mathematical literature for over a century. Beyond its initial role as a formal analytical tool, Jackson's framework has increasingly found relevant applications in modeling complex physical systems. This historical trajectory has inspired various generalizations; however, introducing deformed mathematical structures requires careful consideration to ensure that the fundamental physical properties and conservation laws of the system remain intact.
		
		In the context of $q$-algebraic deformations, the parameter $q$ acts physically as a measure of disorder, non-extensivity, or perturbation. In the non-deformed limit $q \to 1$, the canonical classical and quantum mechanics frameworks are rigorously recovered. The development of alternative $q$-deformed operators offers distinct strategies for incorporating and tuning this deformation parameter. The $q$-derivative proposed in this work provides a new mathematical tool that stems directly from the foundational asymmetric $q$-number bracket. Building upon Heisenberg's operator algebra, this operator maintains structural compatibility with the established Jackson framework while offering an alternative operational form that, to the best of our knowledge, has not been previously reported in the literature.
		
		The symmetric realization of Biedenharn--Macfarlane, discussed in Section \ref{bm}, reinforces this perspective from a complementary viewpoint. Based on a distinct $q$-number convention, the Biedenharn--Macfarlane formulation demonstrates explicitly---through the anharmonic spectrum of the $q$-deformed oscillator and the modified state geometry for systems with dimension $d \ge 3$---how the underlying choice of the $q$-bracket governs observable physical consequences, even as an isolated two-level qubit remains insensitive to this particular deformation. This contrast illustrates a key conceptual point: although different $q$-number constructions share the same classical limit $q \to 1$, they encode distinct physical behaviors when $q \neq 1$. This distinction underscores the importance of systematically investigating new $q$-deformed differential operators grounded in Jackson's formulation, as developed in this study.
				
				\section*{Author Contribuitions}
				
				All authors contributed equally to all stages of the manuscript.

				

			\end{document}